\documentclass[proceedings]{JHEP3}                 

\PrHEP{PrHEP hep2001}
\conference{International Europhysics Conference on HEP}

\usepackage{epsfig}

\newcommand{\beqas}{\begin{eqnarray*}}
\newcommand{\eeqas}{\end{eqnarray*}}
\newcommand{\bit} {\begin{itemize}}
\newcommand{\eit} {\end{itemize}}
\newcommand{\ben} {\begin{enumerate}}
\newcommand{\een} {\end{enumerate}}
\newcommand{\bdi}{\begin{displaymath}}
\newcommand{\edi}{\end{displaymath}}
\newcommand{\bfi}{\begin{figure}}
\newcommand{\efi}{\end{figure}}
\newcommand{\beq}{\begin{equation}}
\newcommand{\eeq}{\end{equation}}
\newcommand{\beqa}{\begin{eqnarray}}
\newcommand{\eeqa}{\end{eqnarray}}

\newcommand{\rmd}{{\rm d}}

\newcommand{\CS} {Chern--Si\-mons}

\newcommand {\wrt}    {with respect to}
\newcommand {\bcs}    {boundary conditions}

\newcommand {\YM}     {Yang--Mills}
\newcommand {\YMth}   {Yang--Mills theory}

\newcommand {\lhs}    {left-hand side}
\newcommand {\rhs}    {right-hand side}
\newcommand {\SM}     {Standard Model}
\newcommand {\ew}     {electroweak}
\newcommand {\ewsm}   {electroweak Standard Model}
\newcommand {\dEdt}   {\frac{{\mathrm d}E}{{\mathrm d}t}}
\newcommand {\diffeq} {differential equation}
\newcommand {\diffeqs}{differential equations}

\title{New analytic results for electroweak baryon number violation}

\author{\speaker{Frans R. Klinkhamer}\\
        Institut f\"ur Theoretische Physik, Universit\"at
        Karlsruhe, D--76128 Karlsruhe, Germany\\
        E-mail: \email{frans.klinkhamer@physik.uni-karlsruhe.de}}
\author{Yong-Joong Lee\\
        Institut f\"ur Theoretische Physik, Universit\"at
        Karlsruhe, D--76128 Karlsruhe, Germany\\
        E-mail: \email{lee@particle.physik.uni-karlsruhe.de}}

\abstract{Real-time anomalous fermion number violation has been investigated for
massless chiral fermions in spherically symmetric $SU(2)$ Yang-Mills gauge
field backgrounds which can be weakly dissipative or even nondissipative.
Restricting consideration to spherically symmetric fermion fields,
a relation has been found between the spectral flow
of the Dirac Hamiltonian
and two characteristics of the background gauge field.
This new result may be relevant to electroweak
baryon number violation in the early universe.}

\begin{document}

\section{Introduction}

The \ewsm~displays an anomalous violation of baryon ($B$) and
lepton ($L$) number \cite{ABJ69,H76}:
\beq \label{DeltaBLlowE}
\Delta B  =  \Delta L  = N_\mathrm{fam} \:\Delta N_\mathrm{CS}\;,
\eeq
with $N_\mathrm{fam}$ the number of families and $\Delta N_\mathrm{CS}$
the change of Chern-Simons number of the $SU(2)$ \YM~gauge fields.
Strictly speaking, this relation holds only for transitions from
\mbox{(near-)}vacuum to (near-)vacuum.
The total energy $E$ of the process is then far below the top of the energy
barrier associated with the Sphaleron \cite{KM84a}, that is,
$E << E_\mathrm{\,Sphal} \approx 10\, \mathrm{TeV}$
for the zero-temperature case.

In fact, 't Hooft calculated the \emph{tunneling}
process, using the Euclidean (imaginary-time) path integral formalism.
For finite action gauge field configurations, the topological charge
\beq
Q \equiv \int_{\mathbb{R}^4} \rmd^4 x \; \tilde{q}(x)
  \equiv -\frac{1}{32\,\pi^2}\int_{\mathbb{R}^4} \rmd^4 x \; \epsilon^{KLMN}\,
\mbox{tr}\left(F_{KL}F_{MN}\right)\;
\eeq
takes on integer values,
\vspace*{-.5mm}
\beq
Q[\,A_\mathrm{finite\; action}\,] = \Delta N_\mathrm{CS} \in \mathbb{Z} \;.
\eeq
The reason is that the gauge fields at infinity are pure gauge and
$Q$ corresponds to the winding number of the map
\vspace*{-2mm}
\beq
\left. S^3 \right|_{x^2=\infty} \rightarrow SU(2) \sim S^3 \;,
\eeq
which equals the change of  \CS~number $\Delta N_\mathrm{CS}$.
This integer $Q$ is then identified with the number of
fermions created or annihilated in the tunneling process.

But the topological charge $Q$ is, in general, a \underline{noninteger}
for physical processes in Min\-kow\-ski spacetime.
The question, then, is \underline{what} determines the real-time electroweak fermion
number violation $\Delta (B+L)$,
in particular for the processes of the early universe at
temperatures above the \ew~phase transition
($T > T_c \approx 10^2 \, \mathrm{GeV}$)?

Here, we report on results obtained by
a direct investigation of the Dirac equation for a
restricted set of gauge fields and vanishing Yukawa coupling to the Higgs field.

\section{Main result}

In this talk, we consider  pure $SU(2)$ \YMth~with a single
isodoublet of left-handed fermions.
The starting point is  the eigenvalue equation of the time-dependent Dirac
Hamiltonian:
\vspace*{-.5mm}
\beq \label{Heq}
H(t,\vec{x}\,)\,\Psi (t,\vec{x}\,)=E(t)\,\Psi (t,\vec{x}\,)\; .
\eeq
Then fermion number violation is related to the  \underline{\sc spectral flow}
$\mathcal{F}_H$ of the Dirac Hamiltonian $H$.
The general definition of spectral flow is as follows:
$\mathcal{F}[\,t_f,t_i\,]$ is the number of eigenvalues
crossing zero from below minus the number of eigenvalues crossing
zero from above, for the time interval $[\,t_i,t_f\,]$ with $t_i < t_f$.
Henceforth, we simply write $\mathcal{F}$ for $\mathcal{F}_H$.
See  Fig.1 for an example and Ref. \cite{C80} for further details and references.

For \emph{strongly dissipative} gauge fields, the following result
holds \cite{C80,GH95,K95}:
\beq
\lim_{t_i \rightarrow -\infty} \;\; \lim_{t_f \rightarrow +\infty}\,
{\cal F}[\,t_f,t_i\,]=\Delta N_\mathrm{CS}\; \mbox{,}
\eeq
which corresponds to Eq. (\ref{DeltaBLlowE}) above.

For \emph{weakly- or non-dissipative spherically symmetric} gauge fields,
a careful analysis of the zero-eigenvalue equation (\ref{Heq})
gives \cite{KL01}:
\beq \label{F}
{\cal F}[\,t_f,t_i\,]=\Delta N_{\chi}[\,t_f,t_i\,]+
\Delta N_{\Theta}[\,t_f,t_i\,]\; \mbox{,} \;\;
\eeq
with $\Delta N_{\chi}$ $=$ $\Delta N_\mathrm{CS}$ for near-vacuum fields.
The new contribution $\Delta N_{\Theta}$ is called the ``twist factor''
of the spherically symmetric gauge field,
whereas $\Delta N_{\chi}$ is called the ``winding factor.''
A spherically
symmetric $SU(2)$ gauge field  solution is called \emph{strongly dissipative},
if both the (3+1)-dimensional and (1+1)-dimensional energy densities
approach zero uniformly for large times ($t$ $\rightarrow$ $\pm\infty$),
and \emph{weakly dissipative},
if the (3+1)-dimensional energy density dissipates with time, but not
the (1+1)-dimensional energy density.

The result (\ref{F}) multiplied by $2\,N_\mathrm{fam}$
answers, in part, the question about $\Delta (B+L)$ raised in the
penultimate paragraph of the previous section.
In addition, we expect  that $\Delta (B-L)  = 0$, since all left-handed isodoublets
of the \SM~contribute equally, at least for the  spherically symmetric fields
considered. In the rest of the talk, we will explain the origin of the result
(\ref{F}), which holds for a single left-handed isodoublet,
and discuss an explicit verification for (weakly dissipative)
L\"uscher-Schechter gauge fields \cite{LS77}.

\FIGURE[t]{
\epsfig{file=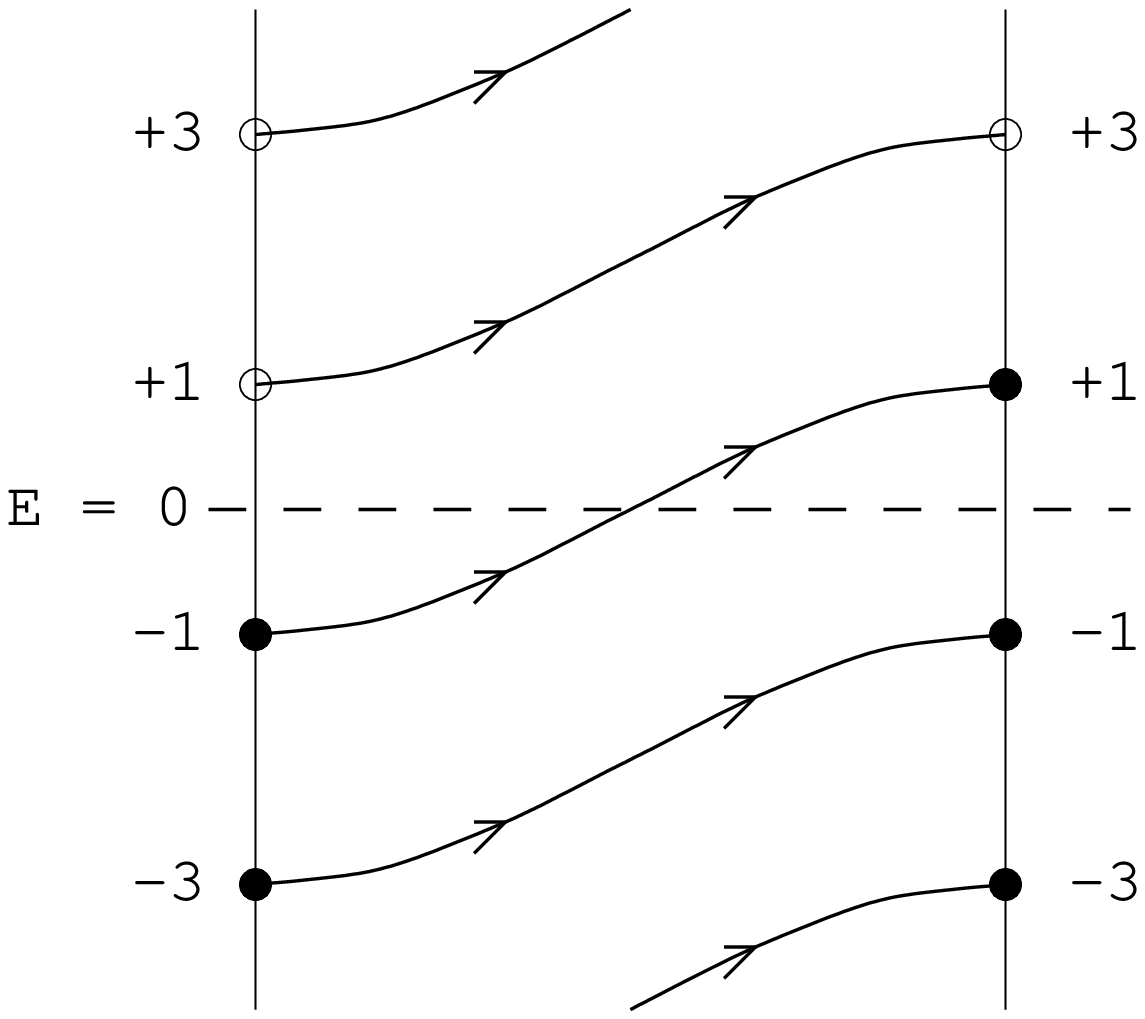,width=0.57\textwidth}
\caption{Sketch of the spectral flow of the  eigenvalues $E(t)$
of the time-dependent Dirac Hamiltonian,
with $\mathcal{F}[\,t_f,t_i\,]=+1$. Filling the (infinite) Dirac
sea at the initial time $t_i$ results in one extra fermion at the final time
$t_f$.}
}

\section{Spherically symmetric \emph{Ansatz}}

The action of chiral $SU(2)$ Yang--Mills theory over Minkowski spacetime (indices $M,N$
running over 0, ... , 3) is given by
\begin{eqnarray}
S     &=& \int_{\mathbb{R}^4} \rmd^{4}x \; \left(
-\frac{1}{2\, g^{2}} \, \mbox{tr} \left(F^{MN}F_{MN}\right)
+ \sum_{f=1}^{N_F}\,\bar{\Psi}_f\,\Gamma^{M}D_{M}\Psi_f\right)\;,
\end{eqnarray}
with the Dirac matrices $\Gamma^{M}$  and the further definitions
\begin{eqnarray*}
F_{MN}&\equiv&\partial_{M}A_{N}-\partial_{N}A_{M} +A_{M} A_{N} - A_{N} A_{M} \;, \quad
A_{M}\equiv A_{M}^{a}\,\sigma^{a}/(2i)\;, \\[2mm]
D_{M} &\equiv& \partial_{M}+A_{M}\,\mbox{P}_{L}\; , \quad
\mbox{P}_{L}\equiv (1-\Gamma_{5})/2\;, \quad
\Gamma_{5}\equiv - i\,\Gamma^{0}\Gamma^{1}\Gamma^{2}\Gamma^{3} \; \mbox{.}\nonumber
\end{eqnarray*}

The spherical \emph{Ansatz} implements
invariance under spatial rotations, modulo $SU(2)$ gauge transformations.
For $N_F=1$, this gives an effective
(1+1)-di\-men\-sional $U(1)$ gauge field theory with gauge fields
$a_0(t,r)$ and $a_1(t,r)$, a complex Higgs-like scalar $\chi(t,r)$ and a
single two-component Dirac spinor $\psi(t,r)$.

\nopagebreak
Explicitly, the effective (1+1)-dimensional $U(1)$ gauge field theory
(indices $\mu,\nu$ running over 0, 1) is given by
\vspace{-1mm}
\begin{eqnarray}
S&=&\left(4\pi/ g^2\right)\;\int_{-\infty}^{+\infty}
\rmd t\int_{0}^{\infty} \rmd r \; \left( \phantom{\frac{1}{1}}\!\!\!\!
\textstyle{\frac{1}{4}}\, r^2f^{\mu\nu}f_{\mu\nu}+|D_{\mu}\chi|^{2}
+ \frac{1}{2}\,(|\chi|^{2}-1)^{2}/r^2  \right. \nonumber \\[2mm]
&& \!\!\!\!\!\! \left. \phantom{\frac{1}{1}}
+ \,g^2\, \bar{\psi} \,[\,\gamma^{\mu}D_{\mu}
+(\mbox{Re}\,\chi+i\gamma_{5}\,\mbox{Im}\,\chi)/r \,]\,\psi \,\right) \;,\label{SAnsatz}
\end{eqnarray}
\vspace{-1mm}\noindent with the definitions
\beqa
f_{\mu\nu}  &\equiv& \partial_{\mu}a_{\nu}-\partial_{\nu}a_{\mu}\;, \quad
D_{\mu}\chi \equiv (\partial_{\mu}-ia_{\mu})\,\chi \;, \quad
D_{\mu}\psi \equiv \left(\partial_{\mu}+i a_{\mu}\gamma_{5}/2\right)\psi\;,
\nonumber\\[2mm]
\gamma^{0}&\equiv& i\sigma^{1}, \quad \gamma^{1}\equiv -\sigma^{3}, \quad
\gamma_{5}\equiv -\gamma^{0}\gamma^{1}\equiv \sigma^{2}\mbox{,} \nonumber
\eeqa
and $\sigma^{a}$, $a=1,2,3$, the $2\times 2$ Pauli spin matrices.
The fermionic part of the action (\ref{SAnsatz}) then gives the precise form of the
Hamiltonian considered in the eigenvalue problem (\ref{Heq}).

\section{Winding and twist factors}

For a  given spherically symmetric $SU(2)$ gauge field at time $t$,
the gauge condition
\beq
\chi (t,0)=\chi (t,\infty )=1\;
\eeq
results in a closed loop in the $\chi$-plane. Now write $\chi(t,r)$ in polar notation:
\beq
\chi (t,r)=\rho (t,r)\,\, \exp\left[\, i\varphi (t,r)\,\right]\,, \quad \rho (t,r) \geq 0 \:.
\eeq
Then define the \underline{\sc winding number} at a fixed time $t$ by
\beq
N_{\chi}(t)\equiv [\,\varphi (t,\infty )-\varphi (t,0)\,]/(2\pi)
\eeq
and the \underline{\sc winding factor} between an initial time $t_i$ and final time
$t_f$ by
\beq \label{windingfactor}
\Delta N_{\chi}[t_{f},t_{i}]\equiv N_{\chi}(t_{f})
-N_{\chi}(t_{i})\,\mbox{.}
\eeq
For near-vacuum fields, it can be verified that
$\Delta N_{\chi} = \Delta N_\mathrm{CS}$.

After a unitary transformation and the elimination of an irrelevant
phase factor, the resulting real Dirac spinor $\tilde{\psi}(t,r)$
can be written as
\beq
\tilde{\psi}(t,r) \equiv |\tilde{\psi}(t,r)|\,
\left(\begin{array}{c}
\sin \Theta(t,r)\\
\cos \Theta(t,r)
\end{array}\right)\mbox{.}
\eeq
The  zero-eigenvalue equation (\ref{Heq}) at fixed $t$ now gives two coupled \diffeqs:
\beq \label{Thetaeq}
\!\partial_{r}\Theta(t,r)         = -\lambda(t,r)\,\sin 2\Theta(t,r) +{\cal R}(t,r)\,,
\quad
\partial_{r}|\tilde{\psi}(t,r)| = |\tilde{\psi}(t,r)|\,\lambda(t,r)\,\cos 2\Theta(t,r) \,,
\eeq
with \bcs
\beq
\Theta (t,0)=0\,, \quad  |\tilde{\psi} (t,0)|=0\;,
\eeq
and definitions
\beq
\lambda(t,r) \equiv \rho(t,r)/r\,, \quad
{\cal R}(t,r)\equiv \left[\, a_{1}(t,r)-\partial_{r}\varphi(t,r) \,\right]/2 \;.
\eeq

It has been shown in Ref. \cite{KL01}  that the first \diffeq~in (\ref{Thetaeq})
can be transformed into a generalized Riccati equation.
For the (unique) solution $\Theta(t,r)$ of this equation,
define the \underline{{\sc spinor twist number}} at a fixed time $t$ by
\beq
N_{\Theta}(t)\equiv [\,\Theta (t,\infty )-\Theta (t,0)\,]/\pi
\;\mbox{,}
\eeq
and the \underline{\sc twist factor}  between an initial time $t_i$ and final time
$t_f$ by
\beq \label{twistfactor}
\Delta N_{\Theta}[t_{f},t_{i}]\equiv N_{\Theta}(t_{f})
-N_{\Theta}(t_{i})\;\mbox{.}
\eeq
Remark that
the twist factor $\Delta N_{\Theta}$ measures an \emph{intrinsic property}
of the spherically symmetric gauge field configuration:
\beq
\Delta N_{\Theta}[t_{f},t_{i}]=\frac{1}{\pi}
\int_{0}^{\infty}\rmd r\int_{t_{i}}^{t_{f}}\rmd t \;
\partial_t \,\partial_r \Theta (t,r)\; ,
\eeq
with $\Theta(t,r)$ an explicitly known functional of the background gauge field,
$\Theta = \Theta[\,\chi,a_1\,]$.
Whether or not there exists a more direct way to
obtain $\Delta N_{\Theta}$ remains an open question.

\section{Relation between spectral flow and gauge field background}

Now consider an arbitrary  fermion zero mode at $t=t^*$.
A straightforward perturbative analysis of the time dependence of the
zero-eigenvalue equation (\ref{Heq}) for the Dirac Hamiltonian gives
\emph{locally}
\beq \label{localF}
\mbox{sgn}\left[\left.\dEdt \right|_{t=t^{*}}\,\right]=
\left.\delta N_{\chi}\right|_{t=t^{*}}+
\left.\delta N_{\Theta}\right|_{t=t^{*}},\; \;
\eeq
with the general definition
\beq
\left. \delta N\right|_{t=t^{*}} \equiv
\lim_{\epsilon\downarrow 0} \left[\, N(t^* +\epsilon) - N(t^* -\epsilon)\,\right] \;.
\eeq
For a finite time interval $[\,t_i,t_f\,]$ with $t_i < t_f$,
this results in the \emph{over-all} spectral flow:
\beq \label{calF}
\!\!{\cal F}[\,t_f,t_i\,]=\Delta N_{\chi}[\,t_f,t_i\,]+\Delta
N_{\Theta}[\,t_f,t_i\,]\; , \;
\eeq
in terms of the winding and twist factors defined by Eqs.
(\ref{windingfactor}) and (\ref{twistfactor}), respectively.
The results (\ref{localF}) and (\ref{calF}) hold for generic gauge field
backgrounds; see Ref.  \cite{KL01}.

The relation  (\ref{calF}) has the form of an \underline{index theorem},
with a property of the fermions on the \lhs~and a characteristic of the gauge fields
on the \rhs; cf. Ref. \cite{C80}.
Next, we turn to an explicit verification of this relation.

\section{Spectral flow for L\"{u}scher-Schechter gauge field backgrounds}

L\"{u}scher and Schechter have independently obtained analytic solutions
for $SU(2)$ gauge field theory, which describe collapsing and re-expanding shells
of energy \cite{LS77}.
In this section, we discuss the  exact fermion zero modes  and the
corresponding spectral flow for some of these gauge field backgrounds.
Throughout, the same (arbitrary) mass scale is used to make the
spacetime coordinates and energy dimensionless.

The starting point for the L\"{u}scher-Schechter  (LS) solutions is
the following spherically symmetric \emph{Ansatz}:
\beq
a_{\mu}         =-q(\tau )\,\partial_{\mu}w \; ,\quad
\mbox{Re}\,\chi = 1+q(\tau )\,\cos^{2}w \; ,\quad
\mbox{Im}\,\chi = \textstyle{\frac{1}{2}}\;q(\tau )\,\sin 2w \; ,
\eeq
in terms of the new coordinates
\beqa
\tau &\equiv& \mbox{sgn}(t)\:
\arccos\left(\frac{1+r^2-t^2}{\sqrt{(1+t^2-r^2)^2+4\,r^2}} \right), \quad
w \equiv \arctan\left(\frac{1-r^2+t^2}{2\,r}\right).
\eeqa

In order to obtain a solution of the full $SU(2)$ \YM~field equations,
the \emph{Ansatz} function $q(\tau)$
must solve the following second-order ordinary differential equation:
\beq
\ddot{q}+2\,q\,(q+1)\,(q+2)=0\;,
\eeq
where a dot indicates a derivative \wrt~$\tau$.
Introducing the energy parameter
\beq
\epsilon \equiv \textstyle{\frac{1}{2}}\, \dot{q}^2+
           \textstyle{\frac{1}{2}}\, q^2(q+2)^2\; ,
\eeq
the explicit solution for $\epsilon>1/2$ is  \cite{LS77}:
\beq
q(\tau ) = -1+ \sqrt{1+\sqrt{2\epsilon}} \;
           \mbox{cn}\!\left[\sqrt[4]{8\epsilon}\,(\tau -\tau_{0})|\,m\right],
\eeq
in terms of the  Jacobi elliptic function $\mbox{cn}[u|m]$ with modulus
$m \equiv \left(1+ \sqrt{2\epsilon}\right)/\left(2\sqrt{2\epsilon}\right) < 1$.
This solution has $\Delta N_\chi \neq 0$. (Note that
$\Delta N_{\chi}$ is always 0 for $\epsilon < 1/2$; cf. Ref. \cite{K95}.)

For moderately large energies, we have verified \cite{KL01}
that the spectral flow is given by
\beq \label{calFlowepsilon}
{\cal F}[ \infty, -\infty \,]=
\Delta N_{\chi}[ \infty, -\infty \,]\;, \quad \mathrm{for}\quad
\epsilon < \epsilon^* \approx 5.37071\;.
\eeq
But for $\epsilon > \epsilon^*$ the situation changes drastically.

Henceforth, we consider the particular LS solution with parameters
\beq
\epsilon =20\,, \quad \tau_0  =- K(m)\,/\,\sqrt[4]{8\epsilon}\, \approx -0.54197\;,
\eeq
and topological charge
\beq \label{Qnonint}
Q \approx -0.13 \;.
\eeq
It turns out that the Higgs-like field $\chi(t,r)$ vanishes at three spacetime points:
\beqa
(t_{-1},r_{-1}) &\approx& (-1.89,2.14)\; ,\quad
(t_{0},r_{0})   =       (0,1)         \; ,\quad
(t_{+1},r_{+1}) \approx (+1.89,2.14)\; .
\eeqa
For the corresponding time slices $t=t_{-1}$, $t_0$, $t_{+1}$, there are exact
fermion zero modes; see Figs. 3 and 6 of Ref. \cite{KL01}.
Surprisingly, there are also fermion zero modes at the time slices
\beq
t=\pm \,t_a \approx \pm \,2.92 \; ,
\eeq
see Fig. 7 of Ref. \cite{KL01}.

\nopagebreak
Altogether, there are five fermion zero modes.
A direct calculation gives for the slopes:
\beqa
\left. \dEdt \right|_{t=-t_{a}}&=&
\left. \dEdt \right|_{t=+t_{a}}  \approx
-0.03<0\;,\nonumber\\[0.2cm]
\left. \dEdt \right|_{t=t_{-1}}&=&
\left. \dEdt \right|_{t=t_{+1}}\:\approx
+0.08>0\;, \quad
\left. \dEdt \right|_{t=t_{0}}\approx
-5.00<0\;.
\eeqa
From these level crossings, we have for the spectral flow
(starting from $t=-t_{a}$ and ending at $t=+t_{a}$):
\beq \label{Fdirect}
{\cal F}[ \infty, -\infty \,]=-1+1-1+1-1=-1 \, .\;
\eeq
On the other hand, the gauge field background has
\beq \label{Findirect}
\Delta N_{\chi}[ \infty, -\infty \,]   = +1 \;,\quad
\Delta N_{\Theta}[ \infty, -\infty \,] = -2 \;.
\eeq
Equations (\ref{Fdirect}) and (\ref{Findirect}) together verify our
index theorem (\ref{calF}):
\beq \label{calFepsilon}
{\cal F}[ \infty, -\infty \,]=
\Delta N_{\chi}[ \infty, -\infty \,] + \Delta N_{\Theta}[ \infty, -\infty \,] \;.
\eeq
This result is a direct generalization of Eq. (\ref{calFlowepsilon}),
since $ \Delta N_{\Theta}=0$ for $ \epsilon < \epsilon^*$.
As mentioned in the Introduction, the anomalous change of fermion number,
$\Delta (B+L)$, is
necessarily  an integer and this is indeed the case for the result (\ref{calFepsilon}),
whereas the corresponding topological charge $Q$
from Eq. (\ref{Qnonint}) is definitely a noninteger.

\section{Conclusions}

Restricting to spherically symmetric fields, we have
established a new relation, Eq. (\ref{calF}), between the
spectral flow of the Dirac Hamiltonian
and two characteristics of the background gauge
fields, the winding and twist factors.

This result holds in particular for weakly dissipative or nondissipative
gauge field backgrounds, as existed in the early universe.
Fundamentally, these new effects appear because of the
long-range behavior of the gauge fields.
Recall that the standard result $\partial_\mu J^\mu_{B+L} \propto \tilde{q}$,
which implies Eq. (\ref{DeltaBLlowE}) of the Introduction,
has been derived \cite{ABJ69,H76} from Feynman perturbation theory, with the
interactions at infinity ``turned off.''

The main outstanding problem now is to understand
the role of the twist factor in the
full (3+1)-dimensional $SU(2)$ Yang--Mills
theory, not just the subspace of spherically symmetric configurations.
Any good idea would be most welcome!

\acknowledgments
We thank the organizers of the
conference and the parallel session for the opportunity to
present our results.

\nopagebreak


\begin{thebibliography}{99}
\bibitem{ABJ69}S.L. Adler, \pr{177}{1969}{2426};
               J.S. Bell and R. Jackiw, \nc{A 51}{1969}{47}.
\bibitem{H76}  G. 't Hooft, \prl{37}{1976}{8}; \prd{14}{1976}{3432}.
\bibitem{KM84a}F.R. Klinkhamer and N.S. Manton, \prd{30}{1984}{2212}.
\bibitem{C80}  N.H. Christ, \prd{21}{1980}{1591}.
\bibitem{GH95} T.M. Gould and S.D.H. Hsu, \npb{446}{1995}{35}.
\bibitem{K95}  V.V. Khoze, \npb{445}{1995}{270}.
\bibitem{KL01} F.R. Klinkhamer and Y.J. Lee,
               \prd{64}{2001}{065024} [\hepth{0104096}].
\bibitem{LS77} M. L\"{u}scher, \plb{70}{1977}{321}; B. Schechter, \prd{16}{1977}{3015}.
\end{thebibliography}
\end{document}